\documentclass{aastex631}

\def\deg{$^\circ$}

\begin{document}

\title{Evidence of a Quasi-periodic Global-scale Oscillation in the
 Near-Surface Shear Layer of the Sun}

\author[0000-0002-0910-459X]{Richard S. Bogart}
\affiliation{W. W. Hansen Experimental Physics Laboratory, Stanford University, Stanford, CA, 94305-4085, USA}

\author[0000-0002-2307-0808]{Charles S. Baldner}
\affiliation{W. W. Hansen Experimental Physics Laboratory, Stanford University, Stanford, CA, 94305-4085, USA}

\author[0000-0002-6163-3472]{Sarbani Basu}
\affiliation{Department of Astronomy, Yale University, PO Box 208101, New Haven, CT 06520-8101, USA}

\author[0000-0002-3834-8585]{Rachel Howe}
\affiliation{School of Physics and Astronomy, University of Birmingham, Edgbaston, Birmingham B15 2TT, UK}

\author[0000-0003-0172-3713]{Maria Cristina Rabello Soares}
\affiliation{W. W. Hansen Experimental Physics Laboratory, Stanford University, Stanford, CA, 94305-4085, USA}
\affiliation{Physics Department, Universidade Federal de Minas Gerais, Belo Horizonte 31270-901, Brazil}

\begin{abstract}
We present evidence of  hitherto undiscovered global-scale oscillations in the near-surface shear layer of the Sun. These oscillations are seen as large scale variations of radial shear in both the zonal and meridional flows relative to their mean values. The variations cover all or most of a visible hemisphere, and reverse with a timescale on the order of a solar rotation. A large annual variation in the meridional shear anomaly is understandable in terms of the tilt of the rotation axis, but the rapid oscillations of the shear anomalies in both zonal and the meridional directions appear to be modulated in a more complex, not-quite annual way, although the latter are also strongly modulated by the projected rotational axis angle. Small-scale anomalies in the neighborhood of active regions lend support to their solar origin and physical interpretation. These results were obtained by analyzing ring-diagram fits of low-order modes in high-resolution Doppler data from the Helioseismic and Magnetic Imager on the Solar Dynamics Observatory.
\end{abstract}

\section{Introduction}
The near-surface shear layer of the Sun is a zone extending for about 35 Mm below the photosphere, or about 5\% of the radius, 0.05 $R_0$, depending somewhat on latitude, in which the local mean differential rotation rate rapidly increases with depth \citep[][etc.]{foukal, rhodes1990, mjt1996, schou1998}. In a uniformly rotating coordinate frame, the variation across the layer can be thought of as a vertical (radial) shear in a mean azimuthally-symmetric zonal flow characteristic of the latitude (and epoch). The depth of the layer and the amplitude of the shear are thought to play a significant role in the evolution of global magnetic activity and the life of magnetic active regions \citep{brandenburg, pipin, karak2016, jha21}. There is also a mean meridional circulation in each hemisphere at the surface, whose associated flows extend downward through and probably beyond the near-surface shear layer \citep[][etc.]{irene,Giles}. Relative to these mean or slowly varying structures, however, there also appear to be anomalous localized flows persisting for comparatively short times. Evidence for slight prograde and retrograde anomalies in the zonal flow patterns in the near-surface regions around certain longitudes at high latitudes, persisting for a few rotations, has been previously presented \citep{Hathaway,poles}. In further exploring these localized anomalous flows, we have discovered what we believe to be a hitherto unsuspected phenomenon: occasional slight enhancements or diminutions of both the mean zonal shear, and of any mean meridional shear characteristic of the latitude as well. These anomalies extend over all or most latitudes in large longitudinal sectors, up to the width of a hemisphere or more, persisting for at least several days and for up to several months. Furthermore, the boundaries between these sectors of positive and negative anomalous shear are typically very sharp, spanning no more than about 20 degrees in longitude.

\section{Data Analysis}
The anomalous flow determinations are based on ring-diagram analysis \citep{hill1988} of Dopplergrams from the Helioseismic and Magnetic Imager \citep[HMI;][]{hmi} on the Solar Dynamics Observatory (SDO). The data are analyzed in regions of diameter five heliographic degrees tiling almost the entire disc, with the tiles spaced every 2.5 degrees in latitude and every 2.5 degrees in longitude as well, at least up to latitude $\pm 40$\deg, with wider spacing at higher latitudes to approximately preserve physical spacing. They are tracked for 9$^h$36$^m$, a little more than their average displacement by rotation of 5\deg\ \citep{Bogart_pipeline}. The analysis is repeated up to 72 times per Carrington rotation, being skipped only when more than 30\% of the potentially available Dopplergrams in the interval are either missing or of unacceptable quality which seldom occurs more than once per rotation.

Although the aim of ring-diagram analysis is to use information from trapped acoustic waves of different radial orders to establish the depth profile of the transverse fluid motions by inverting the data against a profile of the sound speed, the comparatively small extent of the 5\deg\ tiles limits the detection to very short wavelengths; consequently only the very low orders of acoustic modes, up to radial order 3, are regularly fit in the power spectra along with the surface waves, the $f$-mode. Such small data sets are not amenable to traditional inversion techniques, so we instead simply analyze the data for the flow parameters $U_x$ and $U_y$ (representing the zonal and meridional components respectively of the transverse flow field causing a Doppler shift in the wave power) as functions of the equivalent lower turning point corresponding to the phase speed of the mode.

\begin{figure}[ht]
\centering
\includegraphics[width=0.7\textwidth]{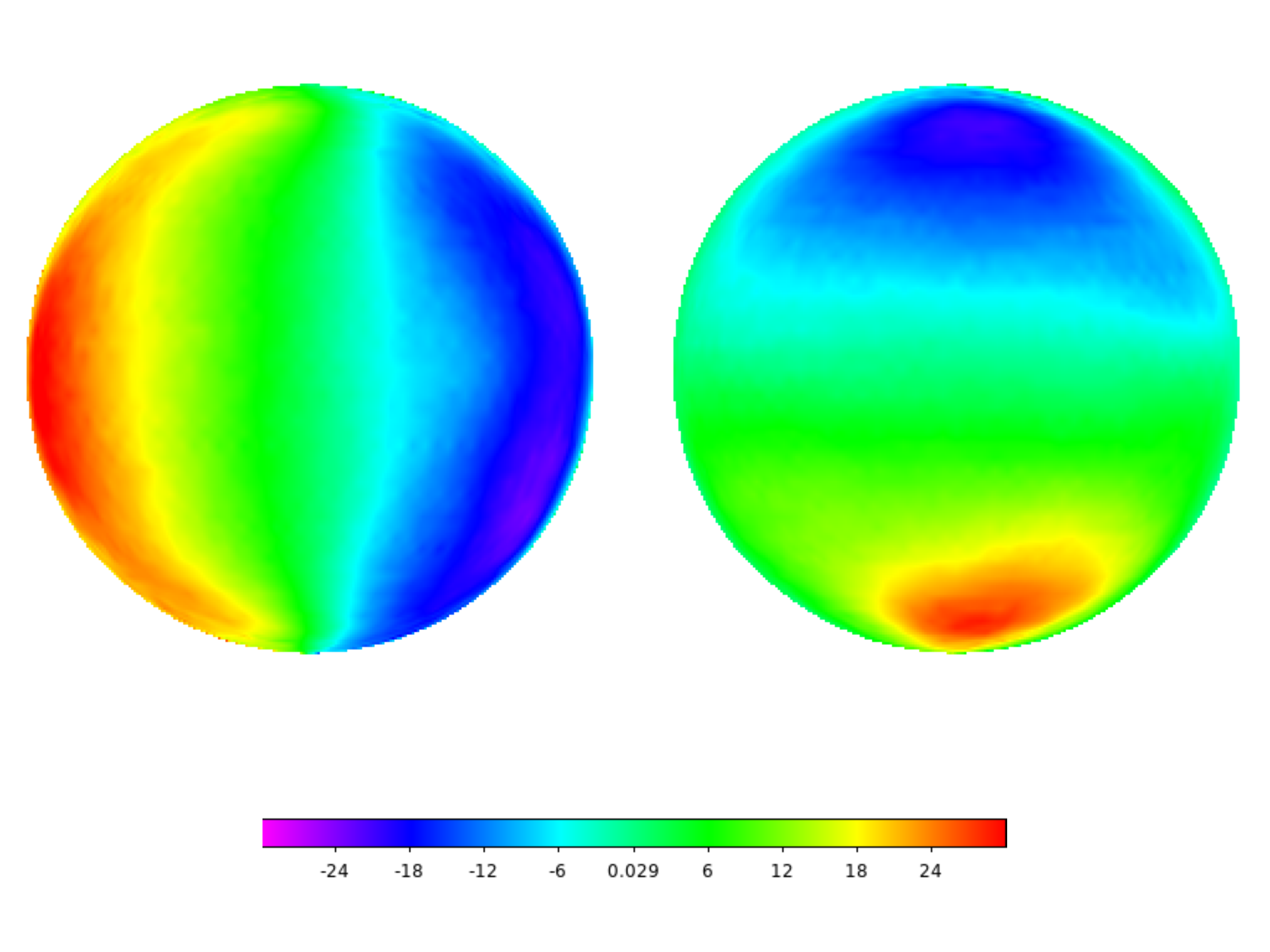}
\includegraphics[width=0.7\textwidth]{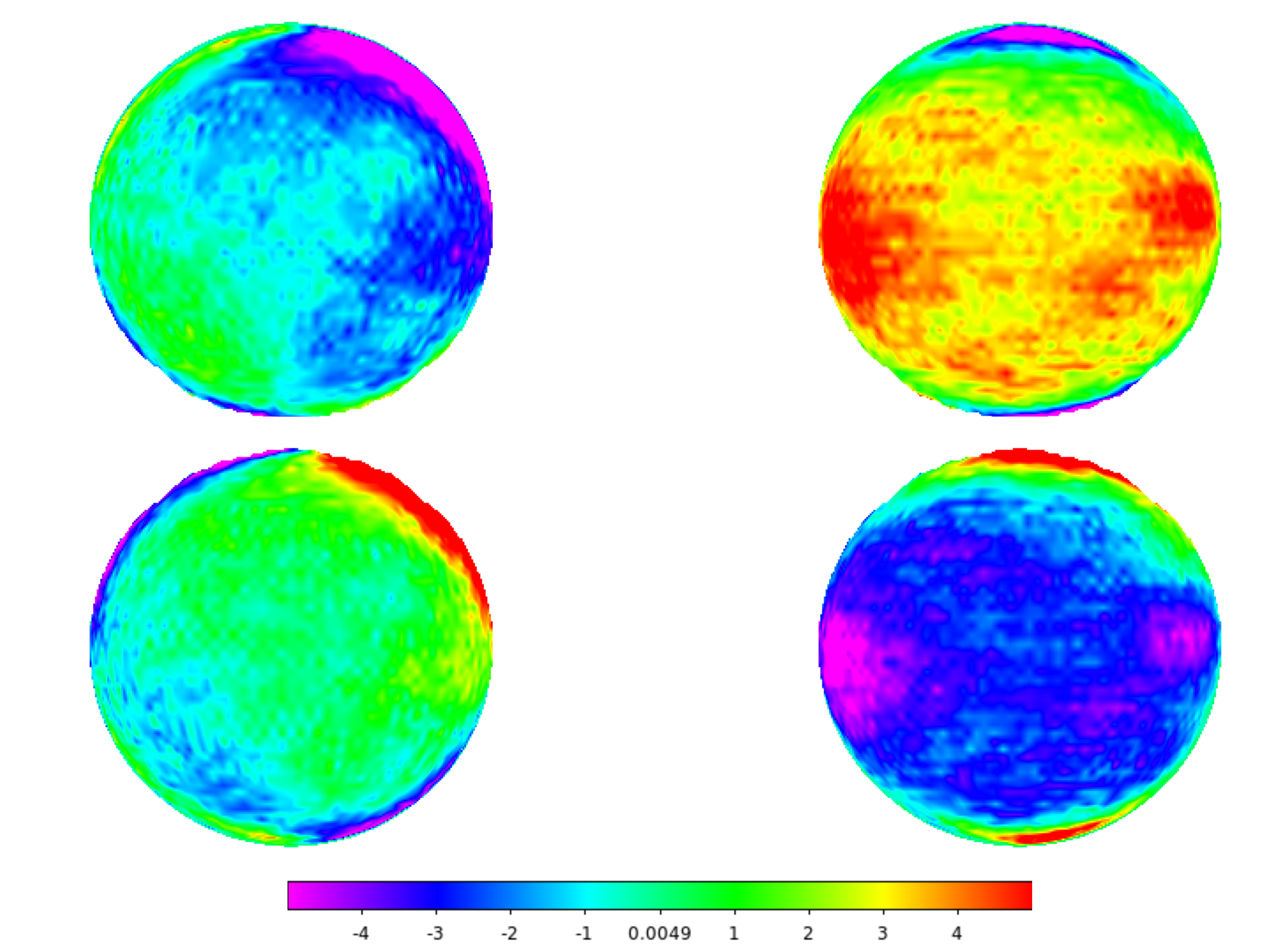}
\caption{
(Top) Annual mean values of the inferred ``shear'' in the zonal and meridional components of the flow, $\langle dU_x/dr\rangle$ (left) and $\langle dU_y/dr\rangle$ (right) respectively at each 5\deg analysis target location, as determined from fitting of the {\sl f-}mode flow parameters between radial turning point values of 0.9950 and 0.9995. (Bottom) Differences in the mean values $\langle dU_x/dr\rangle$ (left) and $\langle dU_y/dr\rangle$ (right) during the 4 months when $B_0 > 3$\deg$.625$ (above) and when $B_0 < -3$\deg$.625$ (below) compared with the annual means. Note the different color scale ranges, which are both in units of m/sec/Mm.
}\label{fig:duidrfmeansyr9}
\end{figure}

For each tile at each time interval, ring-diagram fits produce for each radial order values of the local flow parameters $U_x$ and $U_y$ and equivalent spherical harmonic degree (corresponding to wave-number) as functions of frequency (module {\bf rdfitc} of the HMI processing pipeline \citep {Bogart_pipeline}). We determine linear fits of the flow parameters as functions of the classical turning point, the depth in the Sun at which the sound speed corresponds to the phase speed for the frequency and wave-number of the mode. The slope of the fit, $dU_i/dR_t$, we take as a proxy for the physical radial shear of the flow in the corresponding direction, and label it as such, $dU_i/dr$. The results presented here are based on fits of the $f$-mode, which show the clearest effects. Strictly speaking there is no turning point based on refraction for such waves. Their effective depth of penetration, however, follows the same trend in sound-speed, and the results we have obtained for the low-order $p$-modes $n$ 1--3 covering a depth range of about 2--9 Mm, are in general agreement with $f$-mode results. There are large-scale variations over the field in both the $U_i$ and $dU_i/dr$ fits that depend on the image location of each tile. For the $U_x$ and $U_y$ parameters, these are clearly chiefly solar in origin, being due to the mean differential rotation and meridional circulation respectively. For the $dU_i/dr$ parameters, these appear to be due primarily to observational and analysis effects, likely involving varying sensitivity of the Doppler signal during the tracking of the tiles. The mean values at each location over a full year are shown in the top panel of Fig. \ref{fig:duidrfmeansyr9}. The reference year chosen is Year IX of the SDO mission, Carrington Time (CT) 2216:060--2230:285 (May 1, 2019 -- Apr 30, 2020), centered around the time of solar minimum. The predominantly east-west variation in the zonal shear measurement indicates that differential rotation is not a factor, even though the data have been uniformly tracked at the Carrington rate, while the predominantly north-south variation in the meridional shear measurement is suggestive of a center-to-limb effect, although neither is quite symmetric. It is important to note that although we are using the annual means at each observing location as our nominal means, there are significant variations over the course of a year in the $dU_y/dr$ values, though not the $dU_x/dr$ values, presumably because of the latitudinal dependence of the former. This is exhibited in the lower panel of Fig. \ref{fig:duidrfmeansyr9}. As we shall see, it has significant consequences for the detection of anomalies in the meridional shear.

In order to measure localized variations of the flow and shear as functions of time, the residual differences between the fit values for each tile and the corresponding average for its Stonyhurst location are recorded. For each ``extended'' Carrington longitude and latitude (including the associated Carrington rotation, and running backward or forward in time as appropriate), the residual values for all samples are averaged together, weighted by the cosine of the Stonyhurst longitude at the time of the sample. These represent all the useful observations at the particular Carrington location, of which there are typically $\sim$30 for latitudes up to $\pm$40\deg and at least 17 for latitudes up to $\pm$65\deg.

\section{Results}
\begin{figure}[ht]
\centering
\includegraphics[width=0.24\textwidth]{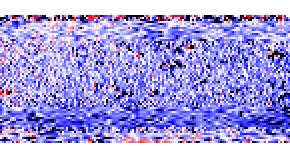}
\includegraphics[width=0.24\textwidth]{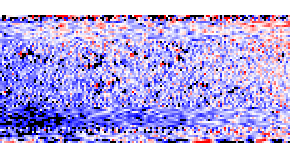}
\includegraphics[width=0.24\textwidth]{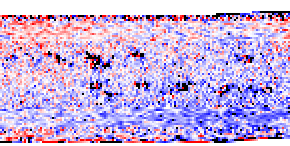}
\includegraphics[width=0.24\textwidth]{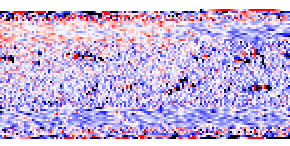}
\includegraphics[width=0.24\textwidth]{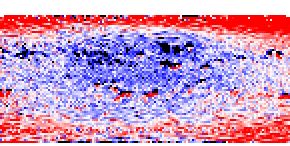}
\includegraphics[width=0.24\textwidth]{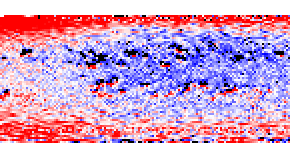}
\includegraphics[width=0.24\textwidth]{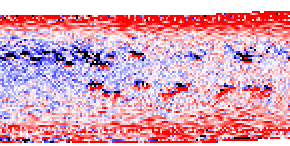}
\includegraphics[width=0.24\textwidth]{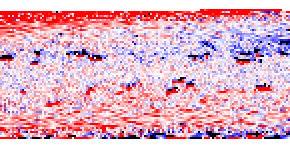}
\caption{
Synoptic maps (in a plate carr\'ee projection with Carrington longitude increasing from left to right and latitude from bottom to top) of the residual values of the zonal shear parameter $dU_x/dr$ (above) and meridional one $dU_y/dr$ (below) for four recent Carrington rotations (2265--2268, Dec. 2022 -- Mar. 2023, right to left in order to align Carrington longitudes, which increase from left to right within each map). The color scale saturation range for both maps is $\pm 10$ m/sec/Mm, with positive (westward and northward) values red. A number of active region sites are conspicuous, including for example those of AR 13190 at 120-15 in CR 2266 and AR 13234 at 345+25 in CR 2268
}\label{fig:synopmapsrecent}
\end{figure}

Plotting the residual values of the presumed radial shear of the flow components at each location and time as synoptic maps, two features can be noted (see Fig. \ref{fig:synopmapsrecent}). One is that in the vicinity of active regions, there is generally convergent shear in both the zonal and meridional components of the flow. This is consistent with both surface and helioseismic determinations that show inflows at the surface and outflows at depth around sunspots \citep[][etc]{haber2003, haber2004, sasha2006}, and lends support to the identification of our measurements with a physical shear.

\begin{figure}[ht]
\centering
\includegraphics[width=\textwidth]{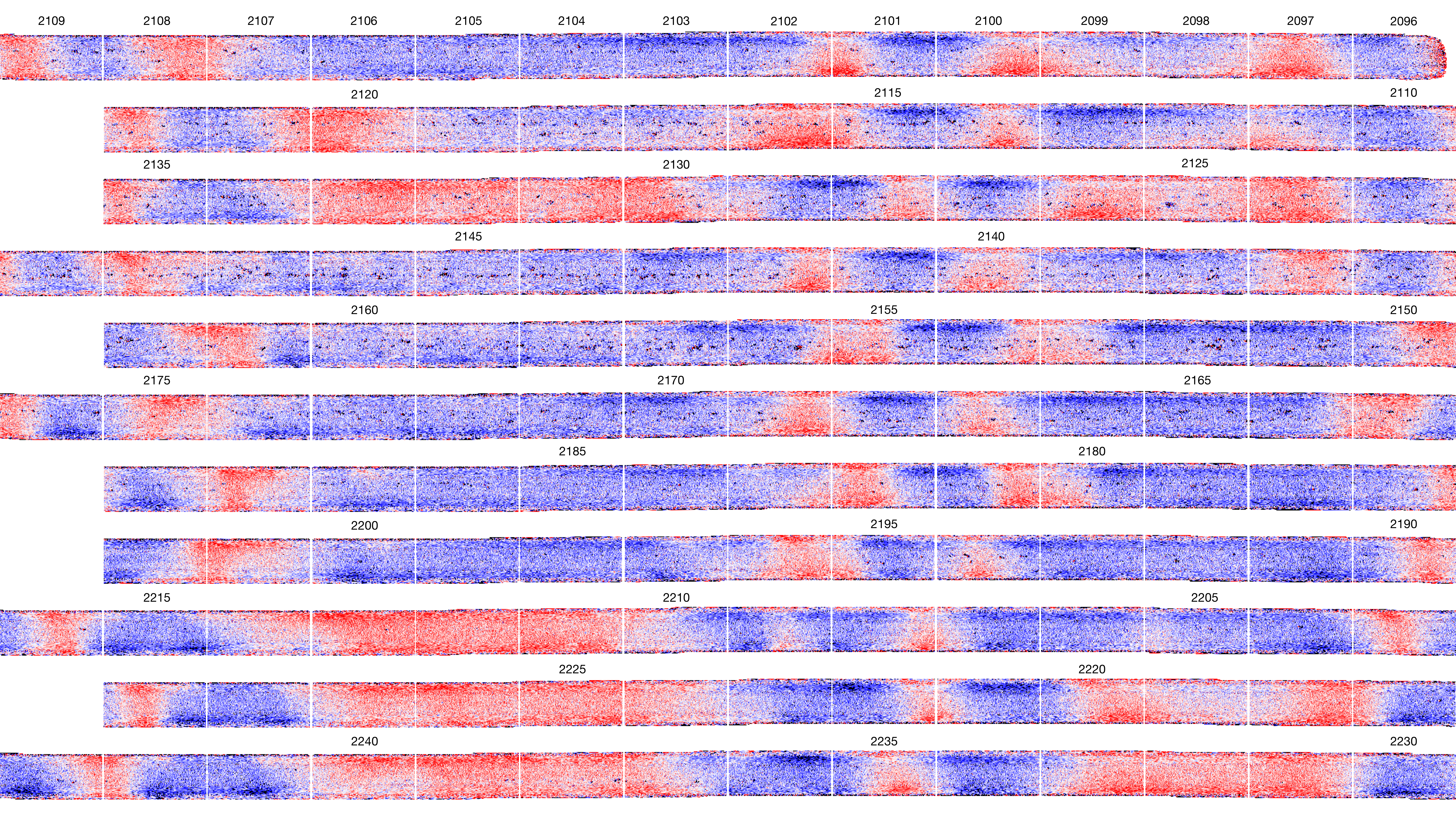}
\includegraphics[angle=-90,width=\textwidth]{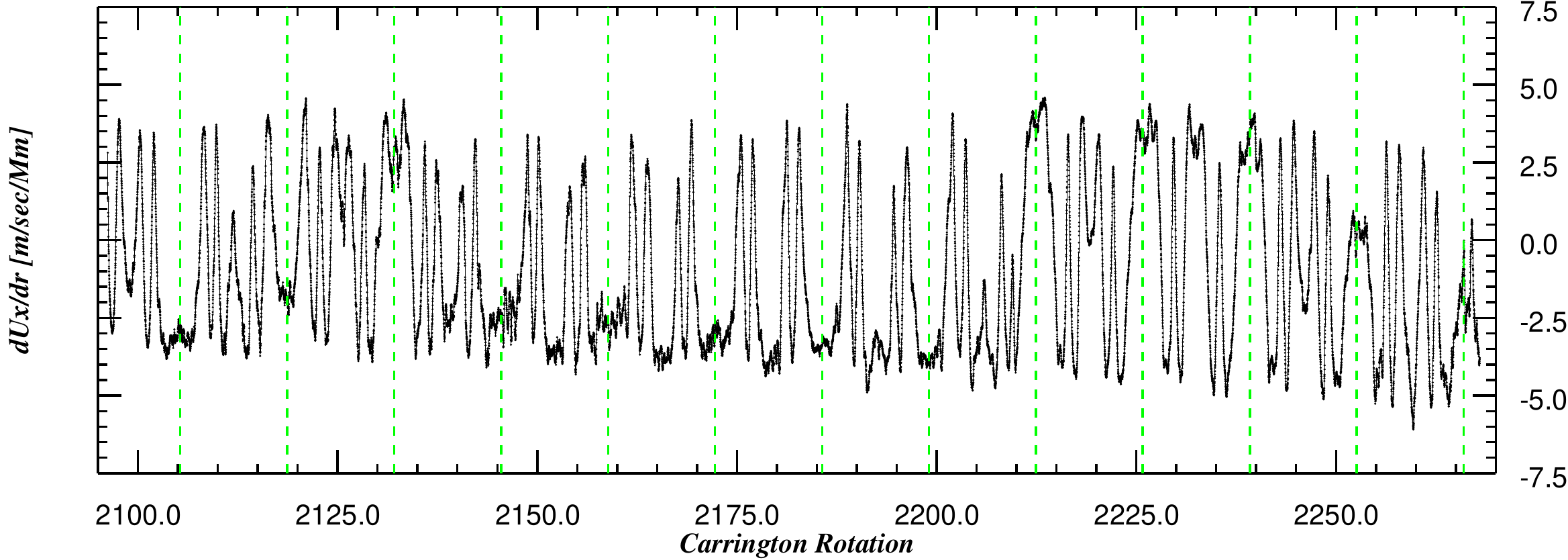}
\caption{
{\sl Top: }Synoptic maps of the zonal shear inferred from $f$-mode binning at each Carrington location for the first eleven years of the SDO mission. Because the Carrington longitudes decrease with time, the plotted rotations increase from right to left within each row, which corresponds approximately to a calendar year. The color scale is the same as for Fig. \ref{fig:synopmapsrecent}. {\sl Bottom:} Time series of 20-sample smoothed averages of the inferred zonal shear at each Carrington longitude, averaged over all latitudes between $\pm40$\deg. The dashed vertical green lines mark the times of the beginning of each calendar year, beginning with 2011.
}\label{fig:duxdrhistory}
\end{figure}

\begin{figure}[ht]
\centering
\includegraphics[width=\textwidth]{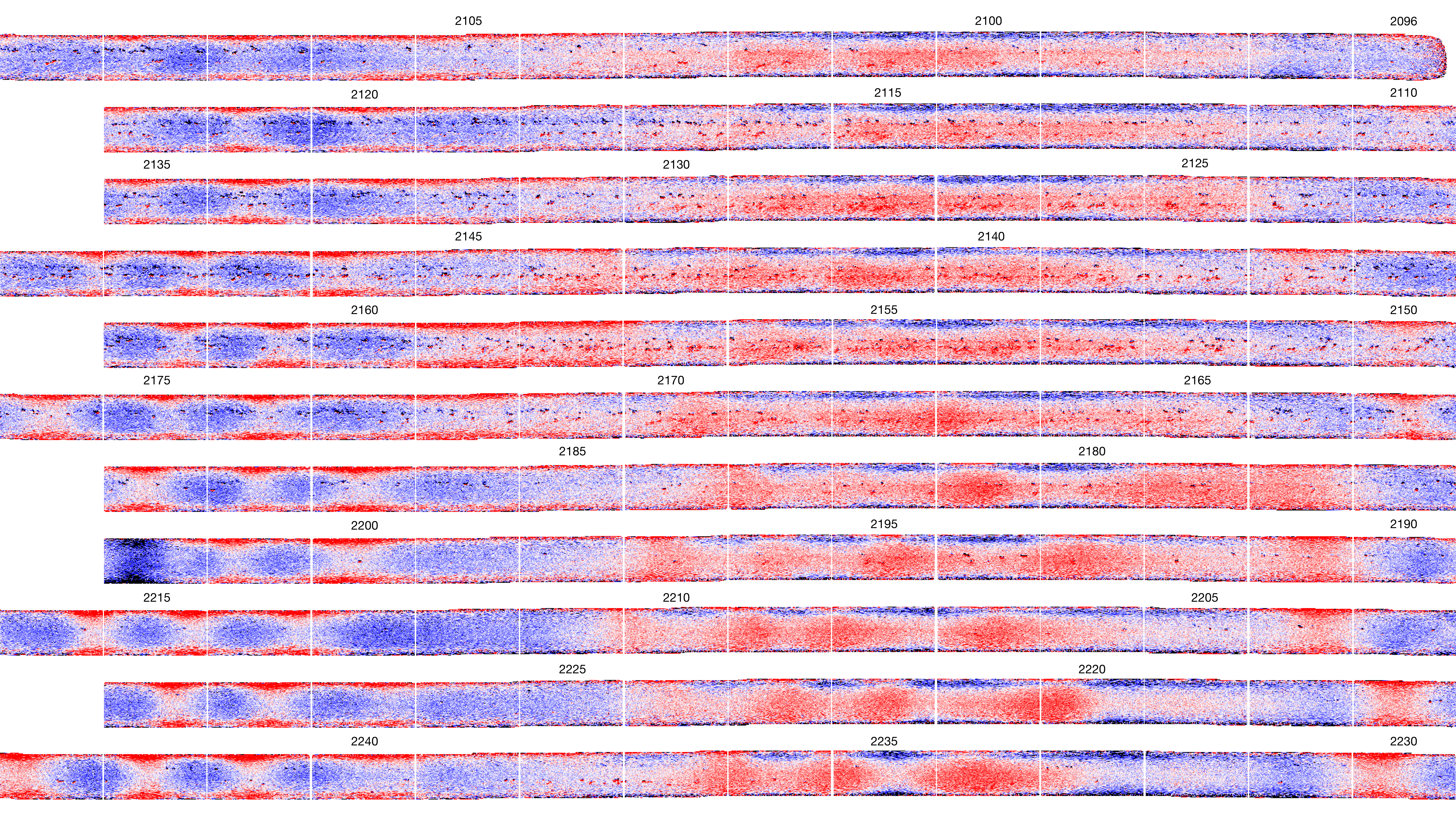}
\includegraphics[angle=-90,width=\textwidth]{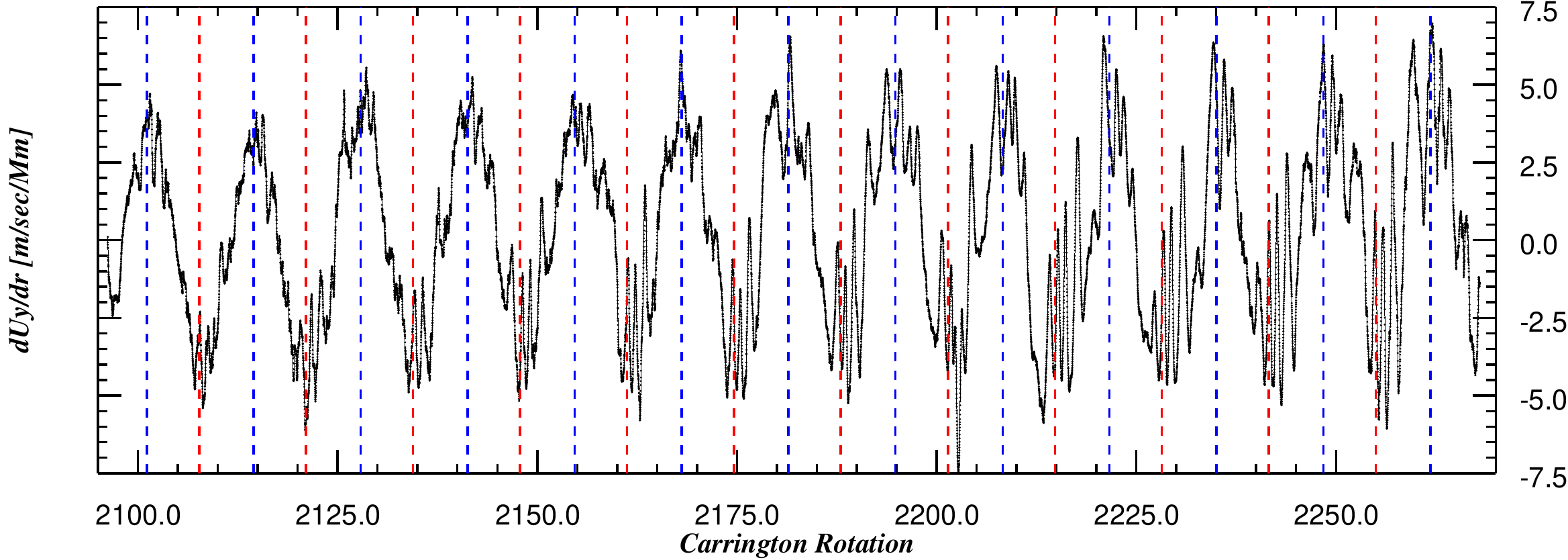}
\includegraphics[angle=-90,width=\textwidth]{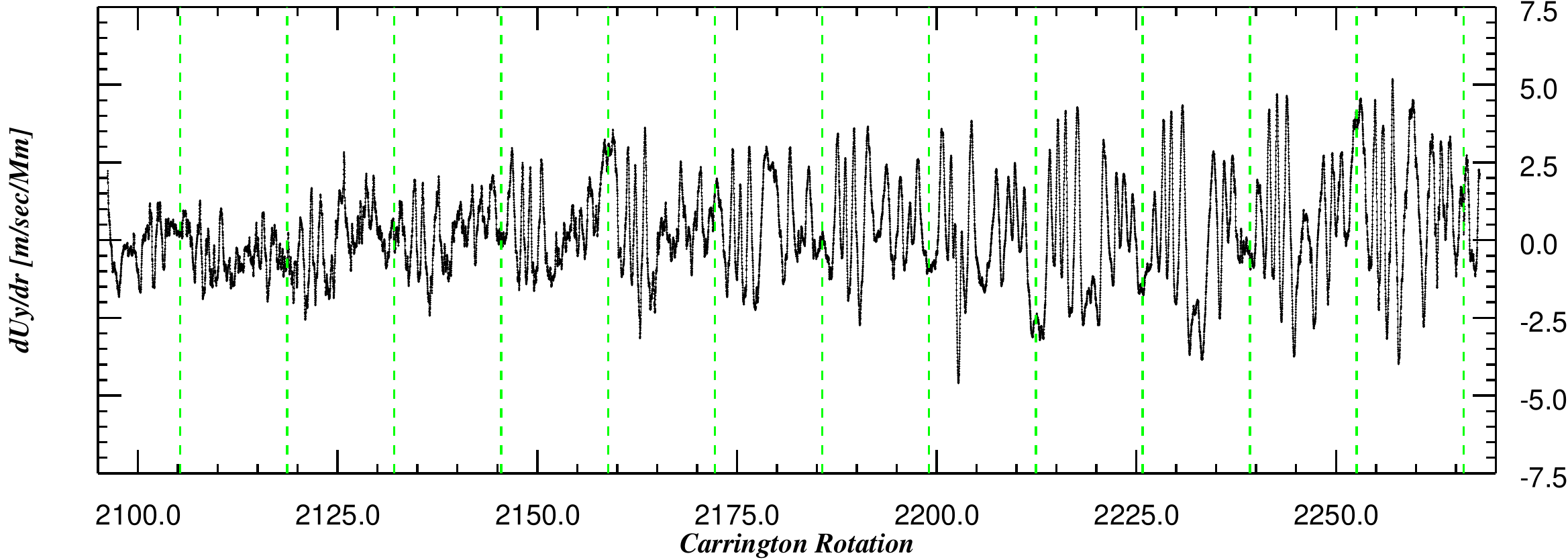}
\caption{
Similar to Fig. \ref{fig:duxdrhistory}, for the meridional shear. The two lower plots are of the time series of the zonally-averaged shear anomalies before (above) and after (below) detrending by a fit for linear dependence on the $B_0$ angle. The dashed blue and red lines in the plot of raw values mark the times of maximum and minimum $B_0$ angle respectively, while the dashed green lines in the plot of detrended values mark the beginnings of calendar years, as in Fig. \ref{fig:duxdrhistory}.
}\label{fig:duydrhistory}
\end{figure}

A more striking and wholly unexpected feature manifests itself however on a larger spatial scale: there are periods, ranging from a few days to a few months, when the residual zonal shear at all or nearly all latitudes is either positive or negative, with rather sharp boundaries between them. The amplitude of this oscillatory behavior, which has persisted throughout the twelve years of the SDO mission to date, is of the order of about $\pm5$ m/s/Mm in the $f$-mode analysis. Similar analyses for the low-order acoustic mode $p$1--3 show an identical pattern, although the amplitude for the $p$1-mode results is somewhat lower than for the others.

Similar patterns exhibit themselves in the meridional shear anomalies as well, although these are superimposed on a strong annual variation that is clearly associated with the projected tilt of the solar rotation axis, the $B_0$ angle. That this should be so is not surprising, as resolution near the limb is very much poorer in the radial than the tangential direction due to foreshortening. At high latitudes, the meridional component is nearly radial, while the zonal component is roughly tangential. (The opposite obtains at low latitudes and extreme longitudes east and west, with the zonal flow and shear measurements being poorer; this is the reason for the weighting by the cosine of the contributing Stonyhurst longitude.) Because the variation of the mean $\langle dU_y/dr\rangle$ over the field is predominantly north-south, it is much more likely to be affected by variation in the $B_0$ angle with time of year than the $\langle dU_x/dr\rangle$ variation, which is basically east-west. This is clearly illustrated in the lower panel of Fig. \ref{fig:duidrfmeansyr9}, in which the means over only the parts of a year for times when the $B_0$ angle is large are compared with the mean for the whole year. At such times these differences are contributing significantly to the local anomalies, which are of necessity with respect to the full-year means at each latitude. Despite this effect, the short-period oscillations continue to be clearly visible, as shown in the middle panel of Fig. \ref{fig:duydrhistory}. Furthermore, after removal of the annual variations, which are well fit by $dU_y/dr = 0.5 B_0$, with $dU_y/dr$ in m/s/Mm and $B_0$ in deg, the residual amplitude of the oscillations in the meridional shear are approximately the same as those of the zonal shear as shown in the bottom panel, although their amplitude has been more or less steadily increasing since the beginning of mission, when they were scarcely distinguishable.

A puzzling feature of the observed oscillations is a nearly (but not quite) annual pattern in their recurrence. It is evident for the zonal shear anomalies, which are not significantly affected by variation of the $B_0$ angle. This can be clearly seen in Fig. \ref{fig:duxdrhistory}. There are typically sets of multiple peaks and troughs recurring on a time scale of about a solar rotation in spring and autumn (when $|B_0|$ is large), while there are also extended periods around midwinter and midsummer (the times of perihelion and aphelion when mean observer radial velocity is at a minimum) when the anomalous shear remains of the same sign. (Note that the SDO mission began around May 1 2010, so the first such period of suppressed oscillation was centered around CR 2104/2105. For reference, the Jan 1 times are marked in the lower panel.) Note however that the zonal shear anomalies during these extended times were of different signs: with one exception they were negative at the beginnings of years 2011--2018, and positive in 2013 and 2019 and thereafter. Comparable patterns can be seen in the meridional shear anomalies after removal of the annual $B_0$ variation: there are periods of rapid oscillations around spring and autumn, and other times when these oscillations are suppressed. To some extent this effect is masked by yet another puzzling feature, a gradual but steady increase in the amplitudes of the meridional shear anomaly amplitudes over the course of the mission, while those of the zonal shear anomalies remain constant.

\begin{figure}[ht]
\centering
\includegraphics[width=0.45\textwidth]{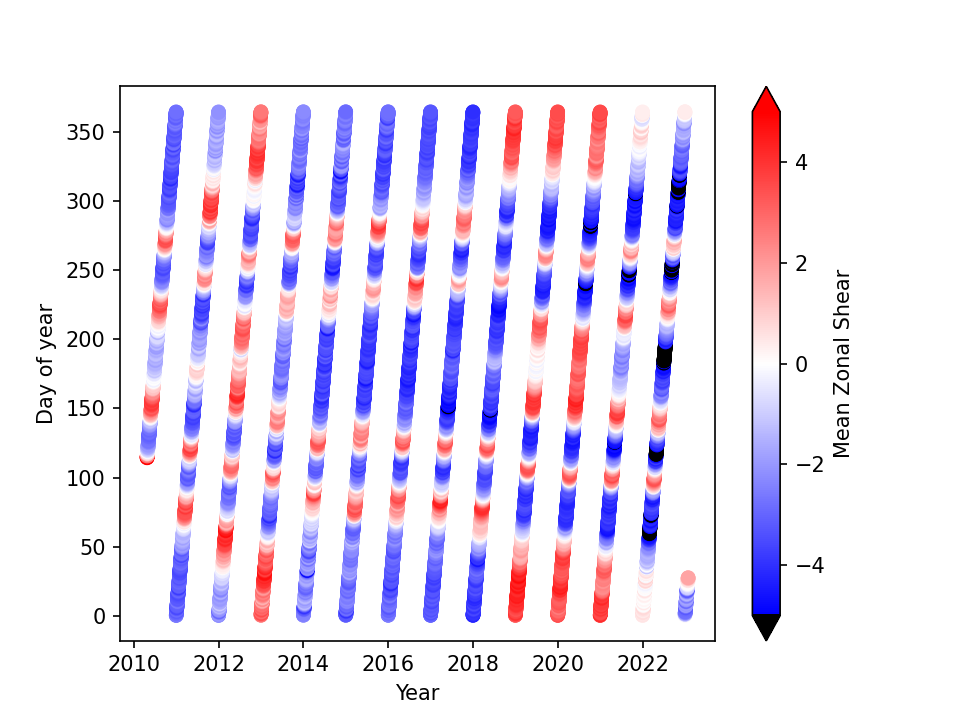}
\includegraphics[width=0.45\textwidth]{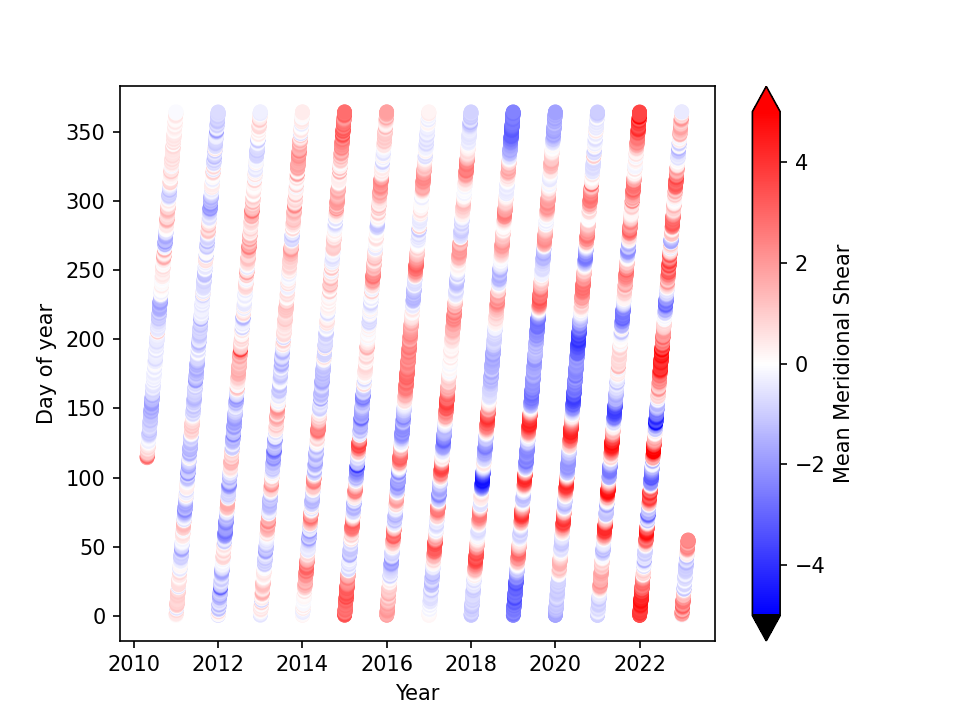}
\caption{``Stack plots'' showing the mean shear anomalies at all latitudes between $\pm$40\deg in the zonal direction (left) and the meridional direction (after detrending by the annual $B_0$ dependence, right), as functions of time on the horizontal axis and day of the year on the vertical axis.}
\label{fig:dofyrstack}
\end{figure}

That the recurrence pattern of the short-term shear oscillations is not quite annual is vividly shown in Fig. \ref{fig:dofyrstack}. From that plot it appears that the date of the spring reversals has been advancing, while that of the autumn ones has been retreating. The change in behavior of the periods when reversals were suppressed is also clear from that figure. Furthermore, the variation in time of the annual phasing of these oscillations is the same {\bf for both the zonal and meridional components}, as shown in Fig. \ref{fig:dofyrstack}, although the discontinuity around the time of solar minimum is much less pronounced {\bf for the meridional component}, indeed if it is present at all. Instead, there seems to have been a gradual increase in the amplitude of these oscillations over the whole observing interval.

\section{Discussion}

Association of changes in the near-surface dynamics with the solar cycle, identified as the torsional oscillation, has been based on azimuthally-averaged measurements of the latitude dependence of the zonal velocity \citep{Howard&LaBonte, Kosovichev&Schou}; likewise for the meridional circulation \citep{Duvall, LaBonte&Howard, Giles}. The distribution of magnetic activity over the photosphere is, however, far from uniform, and it is tempting to associate the sectoral patterns observed here with the development of activity nests and active longitudes \citep{Castenmiller,Balthasar}. It is noteworthy that the zonal shear anomalies, although often exhibiting a north-south hemisphere imbalance in amplitude (clearly associated with the observational effects of the annual $B_0$ variation), nevertheless generally extend over all or nearly all latitudes, well beyond the sunspot zones, and that they persist throughout the solar cycle with no obvious changes in their amplitude nor frequency of occurrence. It should also be remarked however that {\sl(a)} there are long-term variations in the time of year at which rapid oscillations on a timescale of a rotation are seen; and {\sl (b)} the sign of the zonal shear anomaly at least during the periods when it persists for multiple rotations (particularly around the time of perihelion) lasts for several years. There are, however, changes in 2013 and again in 2018. This last of course roughly coincides with the boundary between Solar Cycles 24 and 25, as does the change in phase of the times of observed rapid oscillations. However, it also coincides with the one time during the mission when the focus was adjusted by altering the target temperature of the front window, while the former coincides with the one time when the thermal control program was altered. The remainder of the oscillatory pattern, particularly the sudden alterations of the sign of the anomalies within a solar rotation, cannot be explained by any known instrument or orbital systematics, however; hence we believe that they are solar in nature.

\begin{figure}[ht]
\centering
\includegraphics[width=\textwidth]{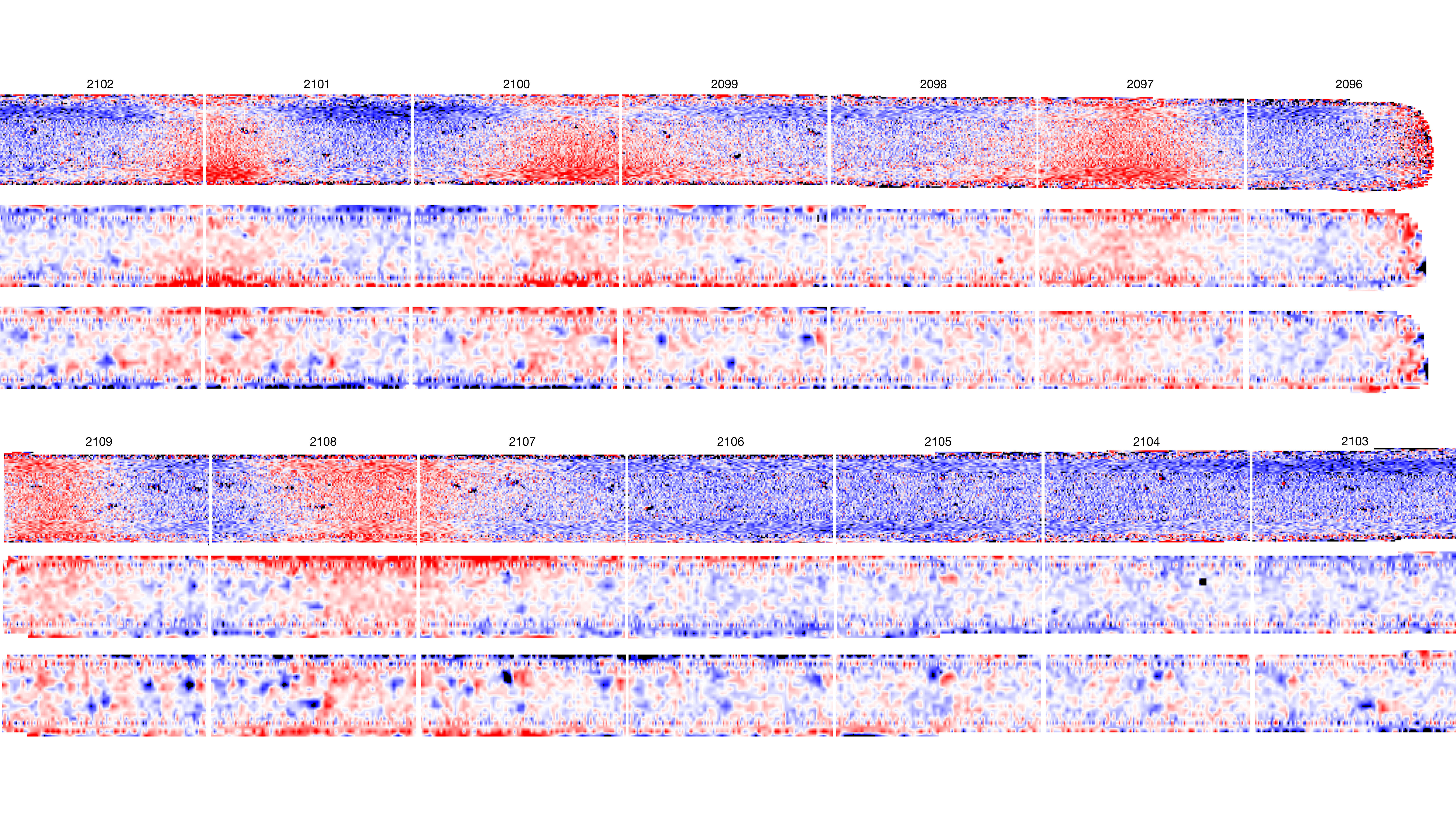}
\caption{
Synoptic maps of the zonal shear inferred from $f$-mode binning of different ring fits at each Carrington location for the first year of the SDO mission. The top rows of each set are for the fits of spectra of 5\deg\ tiles, and together correspond to the top row of the top panel of Fig. \ref{fig:duxdrhistory}. The middle rows of each set are for fits of spectra of 15\deg\ tiles using the same method as for the 5\deg\ tiles ({\bf rdfitc}), while the bottom rows are for fits of the same spectra using the {\bf rdfitf} method. The color scale saturation level for the maps from 5\deg\ tiles is $\pm$10 m/sec/Mm, as in the earlier figures, while it is $\pm$5 m/sec/Mm for the maps from 15\deg\ tiles.
}\label{fig:cf5and15}
\end{figure}

Although the results presented here are based on the fitting of the $f$-mode parameters for transverse flow and wave-number as functions of frequency, the results when fitting the same parameters for the accessible acoustic modes $p$-1--3, as remarked above, are broadly similar in the locations, timings, and sign of the oscillations, though differing somewhat in magnitude. Likewise, although these results are based on analysis of 5\deg\ tiles, they are also supported in analysis of larger tiles as well: synoptic maps similar to those of Figs. \ref{fig:duxdrhistory} and \ref{fig:duydrhistory} based on fitting of 15\deg\ tiles exhibit the same large scale patterns of shifts between positive and negative anomalies, and at the same times, as can be seen in Fig. \ref{fig:cf5and15}. In those maps the localized anomalous shears around active regions are also quite pronounced. The same is true for mode parameters resulting from the alternate fitting method of the 15\deg\ spectra in which they and the mode frequencies are fit as functions of wave-number ({\bf rdfitf}: \citep{Bogart_pipeline} and \citep{rdfitf}), although the results are distinctly noisier. This may be because with the choices of region sizes and tracking times in the HMI pipeline and fitting at steps of $\Delta k$ vs. $\Delta\nu$, that method under-samples the power spectral data, whereas {\bf rdfitc} over-samples the spectra, resulting in substantially more data to which to fit means and derivatives. It suggests the possibility of exploring fittings of the higher order modes accessible in those cases. In this case, the results obtained from fitting of the variation of the flow parameters with turning point depth may be compared with results of actual inversions, although inversions for shear very near the surface are difficult because of the lack of a sufficient number of high-degree modes. We have not detected any large-scale or coherent shear anomalies in the analysis of inversions of the mode parameters resulting from either fitting method applied to 15\deg spectra.

Finally, it should be noted that although there is a clear and simple dependence of the averaged meridional shear anomalies on the annually varying $B_0$ angle, the (not-quite annual) variation in the appearance (or detection) of averaged zonal shear anomalies is much harder to explain. It may lie in the variation of the large SDO velocities, which are known to be associated with substantial diurnal variations in the observed quantities \citep{Couvidat}, and which vary somewhat with precession of the orbital nodes. Likewise, the apparent steady increase in the amplitude of the swings in anomalous meridional shear is puzzling. Whether it may be due to a secular trend in the sensitivity of our detection method, or also associated with a solar cycle, it is too soon to be able to say. The fact that at different times of year the intervals between oscillations either advances or retreats suggests that with the present data set it is unlikely that they can be associated with either a regular prograde or retrogade motion of an azimuthal order $m=1$ disturbance, at least in the Carrington frame. For these reasons, it is highly desirable to confirm these observations using data from other sources.

This work uses data from the Helioseismic and Magnetic Imager. HMI data are courtesy of NASA/SDO and the HMI science team. The data used in this article are publicly available from the Joint Science Operations Center at jsoc.stanford.edu. The fitted parameters of the ring diagram fits used for the current study are available from the corresponding author on reasonable request. This research was supported in part by NASA Contract NAS5-02139 to Stanford University. RH acknowledges the support of the UK Science and Technology Facilities Council (STFC) through grant ST/V000500/1.

\facility{HMI}


\end{document}